\documentclass[12pt]{article}
\usepackage[english,german,french,polish]{babel}
\usepackage[T1]{fontenc}
\usepackage{amsfonts}

\begin{document}

\selectlanguage{english}

\baselineskip 0.75cm
\topmargin -0.6in
\oddsidemargin -0.1in

\let\ni=\noindent

\renewcommand{\thefootnote}{\fnsymbol{footnote}}

\pagestyle {plain}

\setcounter{page}{1}

\pagestyle{empty}

~~~

\begin{flushright}
IFT--03/10
\end{flushright}

{\large\centerline{\bf Is there a dynamical group structure behind}}

{\large\centerline{\bf the bilarge form of neutrino mixing matrix?{\footnote {Work supported in part by the Polish State Committee for Scientific Research (KBN), grant 2 P03B 129 24 (2003--2004).}}}}

\vspace{0.4cm}

{\centerline {\sc Wojciech Kr\'{o}likowski}}

\vspace{0.3cm}

{\centerline {\it Institute of Theoretical Physics, Warsaw University}}

{\centerline {\it Ho\.{z}a 69,~~PL--00--681 Warszawa, ~Poland}}

\vspace{0.3cm}

{\centerline{\bf Abstract}}

\vspace{0.2cm}

We observe that the {\it invariance} of neutrino mixing matrix under the simultaneous 
discrete transformations $\nu_1\,,\, \nu_2 \,,\, \nu_3 \rightarrow -\nu_1 \,,\, -\nu_2 \,,\, \nu_3 $ 
and  $\nu_e, \nu_\mu \,,\, \nu_\tau \rightarrow -\nu_e \,,\, \nu_\tau \,,\, \nu_\mu $ (neutrino "horizontal conjugation") {\it characterizes} (as a sufficient condition for it) the familiar bilarge 
form of neutrino mixing matrix, favored experimentally at present. Thus, the mass neutrinos 
$\nu_1, \nu_2 , \nu_3 $ get a new quantum number, {\it covariant} with respect to their mixings 
into the flavor neutrinos $\nu_e, \nu_\mu , \nu_\tau $ (neutrino "horizontal parity" \,equal to -1,
-1,1, respectively). The "horizontal parity"~turns out to be  embedded in a group structure 
consisting of some Hermitian and real $3\times 3$ matrices $\mu_1, \mu_2 , \mu_3 $ and 
$\varphi_1, \varphi_2 , \varphi_3 $, forming pairs interconnected through neutrino mixings. 
They generate some discrete transformations of mass and flavor neutrinos, respectively, 
in such a way that the group relations $\mu_1 \mu_2 = \mu_3 $ (cyclic) and $\varphi_1 \varphi_2 
= \varphi_3 $ (cyclic) hold, while $\mu_a \mu_b = \mu_b \mu_a $ and $\varphi_a \varphi_b = 
\varphi_b \varphi_a $. Then, for instance, the $\mu_3$ matrix may be chosen equal to the 
"horizontal parity".

\vspace{0.2cm}

\ni PACS numbers: 12.15.Ff , 14.60.Pq , 12.15.Hh .

\vspace{0.6cm}

\ni April 2003  

\vfill\eject

~~~
\pagestyle {plain}

\setcounter{page}{1}

\vspace{0.2cm}

As is well known, the bilarge form of neutrino mixing matrix, 

\begin{equation}
U = \left( \begin{array}{ccc} c_{12} & s_{12} & 0 \\ - \frac{1}{\sqrt2} s_{12} & \frac{1}{\sqrt2} c_{12} & \frac{1}{\sqrt2}  \\ \frac{1}{\sqrt2} s_{12} & -\frac{1}{\sqrt2} c_{12} & \frac{1}{\sqrt2}  \end{array} \right) 
\end{equation}

\ni (where $c_{23} = 1/\sqrt2 = s_{23}$ and $s_{13} = 0$, while $c_{12}$ and $s_{12}$ are estimated to correspond to $\theta_{12} \sim 33^\circ $), is globally consistent with all present neutrino-oscillation experiments for solar $\nu_e$'s and atmospheric $\nu_\mu$'s as well as with the negative result of Chooz experiment (giving $s^2_{13} < 0.03$) [1] and successful KamLAND experiment [2,3-7] both for reactor $\bar{\nu}_e$'s. However, it cannot explain the possible LSND effect [8] for accelerator $\bar{\nu}_\mu$'s (and $\nu_\mu$'s) whose existence is expected to be clarified soon in the MiniBOONE experiment. Its negative result would exclude mixings of active neutrinos with hypothetical light sterile neutrinos [9], leaving us with the minimal mixing unitary transformation 


\begin{equation}
\nu_\alpha  = \sum_i U_{\alpha i}\, \nu_i \;,
\end{equation}

\ni where $\nu_\alpha = \nu_e , \nu_\mu ,\nu_\tau $ and $\nu_i = \nu_1 , \nu_2 , \nu_3 $ represent the flavor and mass active neutrinos, respectively

In the flavor representation, where the mass matrix for charged leptons is diagonal, the neutrino mixing matrix $U = (U_{\alpha i})$ is at the same time the diagonalizing matrix for neutrino effective mass matrix $M = (M_{\alpha \beta})$. Then,

\begin{equation}
M_{\alpha \beta} = \sum_i U_{\alpha i}\, m_i \, U^*_{\beta i} \,.
\end{equation}

\ni In the case of bilarge form (1) of $U$, the formula (3) gives

\vspace{0.3cm}

\begin{eqnarray}
M_{e e}\; & = & \;\,m_1 c^2_{12} + m_2 s^2_{12} \,, \nonumber \\ 
M_{\mu \mu} & = & \;\,M_{\tau \tau} = \frac{1}{2}(m_1 s^2_{12} + m_2 c^2_{12} + m_3) \,, \nonumber \\ 
M_{e \mu} & = & \! -M_{e \tau} = \frac{1}{\sqrt2}(-m_1 + m_2) c_{12} s_{12} \,, \nonumber \\ 
M_{\mu \tau} & = & \;\, \frac{1}{2}(-m_1 s^2_{12} - m_2 c^2_{12} + m_3) \,.
\end{eqnarray}

\ni Here, $M_{\beta \alpha} = M_{\alpha \beta} =M^*_{\alpha \beta}$~. Making use of Eqs. (4) we can write the neutrino effective mass matrix in the form

\begin{eqnarray}
M & = & \;\,\frac{m_1+m_2}{4} \left( \begin{array}{rrr} 2 & 0 & 0 \\ 0 & 1 & -1 \\ 0 & -1 &1 \end{array} \right) + \frac{m_3}{2} \left( \begin{array}{rrr} 0 & 0 & 0 \\ 0 & 1 & 1 \\ 0 & 1 & 1 \end{array}\right) \nonumber \\ & &\!\!\! + \,\frac{m_2-m_1}{4}\left[ c\left( \begin{array}{rrr} -2 & 0 & 0 \\ 0 & 1 & -1 \\ 0 & -1 & 1 \end{array}\right) + \sqrt2\, s \left( \begin{array}{rrr} 0 & 1 & -1 \\ 1 & 0 & 0 \\ -1 & 0 & 0 \end{array}\right) \right] \,,
\end{eqnarray}

\ni where $c \equiv c^2_{12} - s^2_{12} = \cos 2 \theta_{12}$ and $s \equiv 2c_{12} s_{12} = \sin 2 \theta_{12}$. Here, all three terms, proportional to $m_1 + m_2,m_3$ and $m_2 - m_1$, commute (while two terms proportional to $m_2 - m_1$, anticommute). Diagonalizing $M$ given in Eq. (5), we obtain consistently

\begin{eqnarray}
\left( \begin{array}{rrr} m_1 & 0 & 0 \\ 0 & m_2 & 0 \\ 0 & 0 & m_3 \end{array} \right) & = & \;U^\dagger M U  = \frac{m_1 + m_2}{2} \left( \begin{array}{rrr} 1 & 0 & 0 \\ 0 & 1 & 0 \\ 0 & 0 & 0 \end{array} \right) \nonumber \\ & & \!\!\!+ \,m_3 \left( \begin{array}{rrr} 0 & 0 & 0 \\ 0 & 0 & 0 \\ 0 & 0 & 1 \end{array} \right) + \frac{m_2 - m_1}{2} \left( \begin{array}{rrr} -1 & 0 & 0 \\ 0 & 1 & 0 \\ 0 & 0 & 0 \end{array} \right) \,. 
\end{eqnarray}

\ni The present solar and atmospheric experimental estimates are $\Delta m^2_{21} \equiv m^2_2 - m^2_1 \sim 7\times 10^{-5}\;{\rm eV}^2$ and $\Delta m^2_{32} \equiv m^2_3 - m^2_2 \sim 2.5\times 10^{-3}\;{\rm eV}^2$, respectively, when the case of normal hierarchy $ m_1 < m_2 < m_3$ is considered. Note that $M$ gets here the form

\begin{equation}
M =  \left( \begin{array}{rrr} A & D & -D \\ D & B & C \\ -D & C & B \end{array} \right) \,,
\end{equation}

\ni where $A \equiv M_{e e}$, $B \equiv M_{\mu \mu} = M_{\tau \tau}$, $C \equiv M_{\mu \tau}$ and $D \equiv M_{e \mu} = -M_{e \tau}$ are given in Eqs. (4).

The bilarge mixing matrix $U$ presented in Eq. (1) is not bimaximal as $\theta \sim 33^\circ$ and so,

\begin{equation}
c_{12} \sim 0.84 > \frac{1}{\sqrt2} > s_{12} \sim 0.54 \,.
\end{equation}

\ni But, since both values $c_{12}$ and $s_{12}$ are still large and not very distant from $1/\sqrt2 \simeq 0.71$, one may ask the question, if and to what extent the rough approximation $c_{12} \simeq 1/\sqrt2 \simeq s_{12}$ may work, leading through Eq,. (1) to the approximate bimaximal form of the neutrino mixing matrix

\begin{equation}
U \simeq \left( \begin{array}{rrc} \frac{1}{\sqrt2} & \frac{1}{\sqrt2} & 0 \\ -\frac{1}{2} & \frac{1}{2} & \frac{1}{\sqrt2} \\ \frac{1}{2} & -\frac{1}{2} & \frac{1}{\sqrt2}  \end{array} \right) \,.
\end{equation}

\ni It can be easily seen that in the {\it approximation} (9) for $U$ three discrete transformations of mass neutrinos

\begin{eqnarray} 
\nu_1,\nu_2,\nu_3 & \rightarrow & -\nu_2,-\nu_1,-\nu_3 \;, \nonumber \\
\nu_1,\nu_2,\nu_3 & \rightarrow & \;\; \nu_2, \;\;\, \nu_1,-\nu_3 \;, \nonumber \\
\nu_1,\nu_2,\nu_3 & \rightarrow & -\nu_1,-\nu_2,\;\; \nu_3
\end{eqnarray} 

\ni induce through the mixing unitary transformation (2) three following discrete transformations of flavor neutrinos:

\begin{eqnarray} 
\nu_e,\nu_\mu,\nu_\tau & \rightarrow & -\nu_e,-\nu_\tau, -\nu_\mu \;, \nonumber \\
\nu_e,\nu_\mu,\nu_\tau & \rightarrow & \;\; \nu_e,-\nu_\mu, -\nu_\tau \,,\nonumber \\
\nu_e,\nu_\mu,\nu_\tau & \rightarrow & -\nu_e, \;\; \nu_\tau, \;\; \nu_\mu  \;, 
\end{eqnarray} 

\ni respectively [10]. Moreover, the third Eq. (10) induces the third Eq. (11) {\it strictly}, if the exact form of $U$ defined in Eq. (1) is applied in Eq. (2) [10].

Let us denote the Hermitian and real $3\times 3$ matrices realizing the discrete transformations (10) as

\begin{equation}
{\mu}_1 \equiv \left( \begin{array}{rrr} 0 & -1 & 0 \\ -1 & 0 & 0 \\ 0 & 0 & -1 \end{array}\right)  \;,\; {\mu}_2 \equiv \left( \begin{array}{rrr} 0 & 1 & 0 \\ 1 & 0 & 0 \\ 0 & 0 & -1 \end{array}\right) \;,\;{\mu}_3 \equiv \left( \begin{array}{rrr} -1 & 0 & 0 \\ 0 & -1 & 0 \\ 0 & 0 & 1 \end{array}\right) 
\end{equation}

\ni and those realizing the discrete transformations (11) as

\begin{equation}
\varphi_1 \equiv \left( \begin{array}{rrr} -1 & 0 & 0 \\  0 & 0 & -1 \\  0 & -1 & 0 \end{array} \right) \;,\; \varphi_2 \equiv \left( \begin{array}{rrr} 1 & 0 & 0 \\ 0 & -1 & 0 \\ 0 & 0 & -1 \end{array}\right) \;,\; \varphi_3 \equiv \left( \begin{array}{rrr} -1 & 0 & 0 \\ 0 & 0 & 1 \\ 0 & 1 & 0 \end{array}\right) \;. 
\end{equation}

\ni Then, we can readily show that in the {\it approximation} (9) for $U$ the three equivalent relations [10]

\begin{equation}
\varphi_a U \mu_a  = U \;\;{\rm or}\;\;  U\mu_a = \varphi_a U \;{\rm or}\; \varphi_a = U \mu_a U^\dagger 
\end{equation}

\ni hold for any $a = 1,2,3$. Moreover, for $a = 3$ these three relations are valid {\it strictly}, when the exact form of $U$ given in Eq. (1) is used, since then the third Eq. (10) induces {\it strictly} the third Eq. (11). The first relation (14) tells us that the mixing matrix $U$ is {\it invariant}  under the simultaneous discrete transformations (10) and (11) ({\it approximately} for $a = 1,2$ and {\it strictly} for $a = 3$), while the third relation (14) shows that $\mu_a$ matrices are {\it covariant} under the mixing unitary transformation (2), leading to $\varphi_a$ matrices (again {\it approximately} for $a = 1,2$ and {\it strictly} for $a = 3$). In particular, the matrix 

\begin{equation}
P^{(H)} \equiv \mu_3 = \left( \begin{array}{rrr} -1 & 0 & 0 \\ 0 & -1 & 0 \\ 0 & 0 & 1 \end{array}\right) 
 =  e^{i\, 2\pi\, I^{(H)}_2} = e^{i\, \pi \,\lambda_2}
\end{equation}

\ni with 

\begin{equation}
I^{(H)}_2 \equiv \frac{1}{2} \lambda_2 = \frac{1}{2} \left( \begin{array}{rrr} 0 & -i & 0 \\ i &  0 & 0 \\  0 & 0 & 0 \end{array} \right) 
\end{equation}

\ni may be called the "horizontal parity", getting the eigenvalues -1, -1, 1 for the mass neutrinos $\nu_1,\nu_2,\nu_3$, respectively, when the discrete transformation

\begin{equation}
\left( \begin{array}{r} \nu'_1 \\ \nu'_2 \\ \nu'_3 \end{array} \right) = P^{(H)} \left( \begin{array}{r} \nu_1 \\ \nu_2 \\ \nu_3 \end{array} \right) = \left( \begin{array}{r} -\nu_1 \\ -\nu_2 \\ \nu_3 \end{array} \right)
\end{equation}


\ni --- the "horizontal conjugation" --- is performed [10]. According to Eq. (15) this conjugation is equivalent to a rotation by the angle $2\pi $ around 2-axis in the formal 8-dimensional "horizontal space", where $\lambda_1, \ldots, \lambda_8$ are Gell-Mann matrices acting on the triplet $ (\nu_1,\nu_2,\nu_3)^T$ (then $I^{(H)}_2$ is the 2-component of the "horizontal isospin" $\, \vec{I}^{(H)} = \frac{1}{2} \vec{\lambda}$ with $\vec{\lambda} = (\lambda_1,\lambda_2, \lambda_3) $, while $Y^{(H)} = (1/\sqrt3) \lambda_8$ is the "horizontal hypercharge"). In consequence,

\begin{equation}
\left( \begin{array}{r} \nu'_e \\ \nu'_\mu \\ \nu'_\tau \end{array} \right) =U P^{(H)} U^\dagger \left( \begin{array}{r} \nu_e \\ \nu_\mu \\ \nu_\tau \end{array} \right) = \left( \begin{array}{r} -\nu_e \\ \nu_\tau \\ \nu_\mu \end{array} \right)\;,
\end{equation}

\ni where $P^{(H)}\,\!' = U P^{(H)} U^\dagger = \varphi_3$ and so, the "horizontal parity"~is {\it covariant} with respect to neutrino mixings.

From Eqs. (3) and (14) we infer for any $a = 1,2,3$ that

\begin{equation}
\varphi_a M \varphi_a = M\;\;{\rm or}\;\;M \varphi_a = \varphi_a M  
\end{equation}

\ni {\it i.e.}, the effective mass matrix $M$ is {\it invariant} under the discrete transformations (11) ({\it approximately} for $a = 1,2$ if in addition $m_1 \simeq m_2$, and {\it strictly} for $a = 3$). In fact, 

$$
\varphi_a M \varphi_a = \varphi_a U\,{\rm diag}\,(m_1, m_2, m_3)\, U^\dagger \varphi_a = U\, \mu_a \,{\rm diag}\,(m_1, m_2, m_3)\, \mu_a\, U^\dagger \,,
$$

\ni where

$$
\mu_a \;{\rm diag}\,(m_1, m_2, m_3) \mu_a = \left\{\begin{array}{ll}\;{\rm diag}\,(m_2, m_1, m_3) & {\rm for}\;\;a = 1,2  \\ \;{\rm diag}\,(m_1, m_2, m_3) & {\rm for}\;\;a = 3 \end{array} \right. \,.
$$

\ni Thus, $\varphi_a M \varphi_a \simeq M$ for $a = 1,2$ if in addition $ m_1 \simeq m_2$, and  $\varphi_a M \varphi_a = M$ for $a = 3$).

It is worthwhile to point out that the rough approximation $m_1 \simeq m_2$ goes in the direction shown by the experimental situation, where $\Delta m^2_{21} \sim 7\times 10^{-5}\;{\rm eV}^2$ is considerably smaller than $\Delta m^2_{32} \sim 2.5\times 10^{-3}\;{\rm eV}^2$.

Now, it is important to observe that the matrices (12) and (13) satisfy for $a,b = 1,2,3$ the following algebraic relations:

\begin{equation}
\mu_1 \mu_2 = \mu_3\;\,{\rm (cyclic)} \;\,,\;\,\, \mu_a \mu_b = \mu_b \mu_a \,\;\,,\;\,\, \mu^2_a = {\bf 1} \,\;\,,\;\,\, \mu_1+\mu_2+\mu_3 = -{\bf 1} 
\end{equation}

\ni and

\begin{equation}
\varphi_1 \varphi_2 = \varphi_3\;\,{\rm (cyclic)} \;\,,\;\, \varphi_a \varphi_b = \varphi_b \varphi_a \;\,\,,\;\,\, \varphi^2_a= {\bf 1} \;\,,\;\, \varphi_1+\varphi_2+\varphi_3 = -{\bf 1}
\end{equation}

\ni (but $\mu_a \varphi_b \neq \varphi_b \mu_a$, except for $\mu_3 \varphi_2 = \varphi_2 \mu_3$).

It is easy to see that the matrices $\mu_1, \mu_2,\mu_3$ and $\varphi_1, \varphi_2,\varphi_3$ given in Eqs. (12) and (13) can be used as bases for $3\times 3$ symmetric block matrices of the types

\begin{equation}
\left( \begin{array}{rrr} A_1 & B_1 & 0 \\  B_1 & A_1 & 0 \\  0 & 0 & C_1 \end{array} \right) \;\;{\rm and}\;\; \left( \begin{array}{rrr} A_2 & 0 & 0 \\ 0 & B_2 & C_2 \\  0 & C_2 & B_2 \end{array} \right) \;,
\end{equation}

\ni respectively. The sets of such matrices form two Abelian groups with respect to matrix multiplication, if the inverse of their four blocks exists. They are isomorphic, being related through the unitary transformation generated by the bimaximal mixing matrix $U$ given on the rhs of Eq. (9): $U\,\{1\}\,U^\dagger = \{2\}$, where $\{1\}$ and $\{2\}$ symbolize the sets of matrices of the first and second type (22). The group character of these sets is reflected in the group relations $\mu_1 \mu_2 = \mu_3$ (cyclic) and $\varphi_1 \varphi_2 = \varphi_3 $ (cyclic) for their bases, while their isomorphism corresponds to the unitary transformation $\varphi_a = U \mu_a U^\dagger $ between both bases. Of course, these two groups are Abelian subgroups of the group of all $3\times 3$ nonsingular matrices that can be spun by the basis consisting of {\bf 1} and Gell-Mann matrices $\lambda_1, \ldots, \lambda_8$.

In terms of the matrices (12) and (13) the effective mass matrix presented in Eq. (5) can be rewritten as

\begin{equation}
M = \frac{m_1 + m_2}{4} \left({\bf 1} - \varphi_3\right) + \frac{m_3}{2} \left({\bf 1} + \varphi_3 \right) +\frac{m_2 - m_1}{4} \left[c \left(\varphi_1 - \varphi_2 \right) + \sqrt2\, s \left( \lambda_1 - \lambda_4 \right) \right]\,,
\end{equation}

\ni where

\begin{equation}
\lambda_1 - \lambda_4 = \left( \begin{array}{rrr} 0 & 1 & -1 \\ 1 &  0 & 0 \\ -1 & 0 & 0 \end{array} \right) = \frac{1}{2}\left\{\varphi_3,\mu_1 - \mu_2 \right\}\,.
\end{equation}

\ni When $c_{12} \simeq 1/\sqrt2 \simeq s_{12}$, then $c \simeq 0$ and $s \simeq 1$. If $m_1 \simeq m_2$, Eq. (23) gives

\begin{equation}
M \simeq \frac{m_1 + m_2}{4} \left({\bf 1} - \varphi_3\right) + \frac{m_3}{2} \left({\bf 1} + \varphi_3 \right)  \,.
\end{equation}

\ni In this case, $D \simeq 0$ in Eq. (7). Then, approximately, $M$ is a matrix of the second type (22).

One may speculate in connection with the formula (23) that the $3\times 3$ matrices $\varphi_a$ and $\mu_a \;(a = 1,2,3)$, where $\varphi_1 \varphi_2 = \varphi_3$ (cyclic) and $\mu_1 \mu_2 = \mu_3$ (cyclic), can help us to find the desired {\it dynamical variables} solving hopefully the basic problem of fermion masses. In such a case there may appear a more or less instructive analogy with Pauli matrices, where $\sigma_1 \sigma_2 = i \sigma_3 $ (cyclic), which have led to Dirac matrices solving the problem of fermion spins.

The discrete transformations generated by $\varphi_a$ and $\mu_a$ matrices and the related discrete symmetries may play an important role in Nature because of the absence for neutrinos of electromagnetic and strong interactions. Otherwise, these interactions could largely suppress such fragile, discrete horizontal symmetries that, in contrast to the Standard Model gauge interactions, do not treat equally three fermion generations.

Finally, we should like to point out that both sets of algebraic relations (20) and (21) would still hold, if $\varphi_1, \varphi_2, \varphi_3 $ matrices were defined not by Eqs. (13), but through the relations 

\begin{eqnarray}
\varphi_1 \equiv U \mu_1 U^\dagger & = & \left( \begin{array}{rrr} -s & -\frac{1}{\sqrt2} c & \frac{1}{\sqrt2} c \\ -\frac{1}{\sqrt2} c & -\frac{1}{2} (1-s) & -\frac{1}{2} (1+s) \\ \frac{1}{\sqrt2} c & -\frac{1}{2} (1+s) & -\frac{1}{2} (1-s) \end{array} \right) \stackrel{s\rightarrow 1}{\rightarrow} \left( \begin{array}{rrr} -1& 0 & 0 \\ 0 & 0 & -1 \\ 0 & -1 & 0 \end{array} \right) \;, \nonumber \\ & & \nonumber \\
\varphi_2 \equiv U \mu_2 U^\dagger & = & \left( \begin{array}{rrr} s & \frac{1}{\sqrt2} c & -\frac{1}{\sqrt2} c \\ \frac{1}{\sqrt2} c & -\frac{1}{2} (1+s) & -\frac{1}{2} (1-s) \\ -\frac{1}{\sqrt2} c & -\frac{1}{2} (1-s) & -\frac{1}{2} (1+s) \end{array} \right) \stackrel{s\rightarrow 1}{\rightarrow} \left( \begin{array}{rrr} 1& 0 & 0 \\ 0 & -1 & 0 \\ 0 & 0 & -1 \end{array} \right) \;, \nonumber \\  & & \nonumber \\
\varphi_3 \equiv U \mu_3 U^\dagger & = & \left( \begin{array}{rrr} -1& 0 & 0 \\ 0 & 0 & 1 \\ 0 & 1 & 0 \end{array} \right) \;,
\end{eqnarray}

\ni where $U$ was of its exact form (1) and $\mu_1, \mu_2, \mu_3$ matrices were given as before in Eqs. (12). In this case, our relations (14) would be valid {\it strictly also} for $a = 1,2$, not only for $a = 3$ as before in the case of Eqs. (13). Of course, in the limit of $c_{12} \rightarrow 1/\sqrt2 \leftarrow s_{12}$ {\it i.e.}, $c \rightarrow 0$ and $s \rightarrow 1$, Eqs. (26) would tend to Eqs. (13). Note that, generically, the relations (20) and (21) as well as (14) would hold, if $\varphi_a = \tilde{U} \mu_a  \tilde{U}^\dagger$ with $\tilde{U}$ being any $3\times 3$ unitary matrix and $\mu_a$ were given in Eqs. (12) ($a = 1,2,3$). However, in such a case, one  would get Eqs. (26) only for the unitary matrices $\tilde{U}$ equal to $V_\varphi U V_\mu $, where $U$ would have the form (1), while $V_\varphi $ and $V_\mu $ would be any unitary matrix commuting with $\varphi_a$ and $\mu_a$, respectively [say, $V_\varphi = f_\varphi (\varphi_1, \varphi_2, \varphi_3)$ and $V_\mu = f_\mu(\mu_1, \mu_2, \mu_3)$]. Then, 

\begin{equation}
V_\varphi \varphi_a V^\dagger_\varphi = \varphi_a \equiv V_\varphi U V_\mu\, \mu_a \,V^\dagger_\mu U^\dagger V^\dagger_\varphi \;,\; \varphi_a = U V_\mu \mu_a V^\dagger_\mu U^\dagger = U \mu_a U^\dagger \,,
\end{equation}

\ni where

\begin{equation}
[V_\varphi\,,\, \varphi_a ] = 0 \;,\; [V_\mu\,,\, \mu_a ] = 0 \,.
\end{equation}

\ni Thus, in the class of $V_\varphi U V_\mu $ matrices, one might restrict oneself to the $U$ matrix of the form (1), putting $V_\varphi = {\bf 1}$ and $V_\mu = {\bf 1}$. The form (1) of $U$ is a sufficient condition for the invariances $\varphi_a U \mu_a = U$ with $\mu_a $ and $\varphi_a $ given as in Eqs. (12) and (26), while the form $U \rightarrow V_\varphi U V_\mu $ is also their necessary condition. 

In conclusion, we have introduced two Abelian algebras of Hermitian and real $3\times 3$ matrices $\mu_1, \mu_2, \mu_3$ and $\varphi_1, \varphi_2, \varphi_3$ satisfying the group relations $\mu_1 \mu_2 = \mu_3 $ (cyclic) and $ \varphi_1 \varphi_2 = \varphi_3 $ (cyclic) as well as the constraints $\mu_1 + \mu_2 + \mu_3 =-{\bf 1}$ and $ \varphi_1 + \varphi_2 + \varphi_3 = -{\bf 1}$. These two algebras are isomorphic, as being related through the unitary transformation $ \varphi_a = U \mu_a U^\dagger \;(a = 1,2,3) $, where $U$ is the neutrino mixing matrix. Thus, $\mu_a $ are {\it covariant} with respect to neutrino mixings, leading to $\varphi_a $. Such a unitary transformation implies the {\it invariances} $ \varphi_a M \varphi_a = M$ of the neutrino effective mass matrix $M$: strictly for $a = 3$ and, if $m_1 \simeq m_2$, approximately for $a = 1,2$.

The unitary transformation $\varphi_a = U\, \mu_a \,U^\dagger \;(a = 1,2,3)$ is equivalent to the {\it invariances} $\varphi_a\, U\, \mu_a = U$ of the neutrino mixing matrix $U$. With given $\mu_a$ and $\varphi_a$ matrices as in Eqs. (12) and (26), respectively, these invariances {\it characterize} (as a sufficient condition for them) the monomaximal form ($\theta_{23} = 45^\circ$) of the bilarge mixing matrix $U$ that for $\theta_{12} \simeq 45^\circ$ should be approximately bimaximal ($\theta_{12} \sim 33^\circ$ is the actual experimental estimate).

Summarizing, the algebraic properties of $\mu_a$ matrices can be expressed by the relations

\begin{equation}
\{\mu_1, \mu_2\} = 2\mu_3\;\,{\rm (cyclic)} \;\,,\;\,\, [\mu_a, \mu_b] = {\bf 0} \,\;\,,\;\,\, \mu^2_a = {\bf 1} \,\;\,,\;\,\, \mu_1+\mu_2+\mu_3 = -{\bf 1}
\end{equation}

\ni ($a,b = 1,2,3$). The identical relations hold also for $\varphi_a$ matrices equal to $U \mu_a U^\dagger$. We suggest that $ \varphi_a$ and $ \mu_a$ matrices ($ a = 1,2,3$) play the role of {\it dynamical variables} in the problem of neutrino masses (and, hopefully, of other fermion masses). In fact, according to Eqs. (23) and (24) the neutrino effective mass matrix $M$ can be expressed by means of the matrices {\bf 1}, $ \varphi_3$ and $\mu_1, \mu_2$ ({\bf 1} $=  -\varphi_1 - \varphi_2 - \varphi_3 = - \mu_1 - \mu_2 - \mu_3$) and the parameters $m_1,m_2, m_3$ and $s$, the number of the latter should be certainly reduced, say, by the conjecture that $m_1 {\bf :} m_2 {\bf :}  m_3 \simeq m_e {\bf :} m_\mu {\bf :}  m_\tau$ [10]. 

\vfill\eject

~~~~
\vspace{0.5cm}

{\centerline{\bf References}}

\vspace{0.5cm}

{\everypar={\hangindent=0.6truecm}
\parindent=0pt\frenchspacing

{\everypar={\hangindent=0.6truecm}
\parindent=0pt\frenchspacing

~[1]~M. Appolonio {\it et al.} (Chooz Collaboration), {\it Eur. Phys. J.} {\bf C 27}, 331 (2003).

\vspace{0.2cm}

~[2]~K. Eguchi {\it et al.} (KamLAND Collaboration), {\it Phys. Rev. Lett.} {\bf 90}, 021802 (2003).

\vspace{0.2cm}

~[3]~V. Barger and D. Marfatia, {\it Phys. Lett.} {\bf B 555}, 144 (2003).

\vspace{0.2cm}

~[4]~G.L. Fogli {\it et al.}, {\it Phys. Rev.} {\bf D 67}, 073002 (2003).

\vspace{0.2cm}

~[5]~M. Maltoni, T. Schwetz and J.W.F. Valle, {\tt hep--ph/0212129}.

\vspace{0.2cm}

~[6]~A. Bandyopadhyay {\it et al.}, {\it Phys. Lett.} {\bf B 559}, 121 (2003).

\vspace{0.2cm}

~[7]~J.N.~Bahcall, M.C.~Gonzalez--Garcia and C. Pe\~{n}a--Garay, {\tt hep--ph/0212147v2}.

\vspace{0.2cm}

~[8]~G. Mills (LSND Collaboration), {\it Nucl. Phys. Proc. Suppl.} {\bf 91}, 198 (2001).

\vspace{0.2cm}

~[9]~C. Giunti, {\tt hep--ph/0302173}.

\vspace{0.2cm}

[10]~W. Kr\'{o}likowski, {\tt hep--ph/0303148v2}.

\vfill\eject

\end{document}